\documentclass[11pt]{article}%
\usepackage{amsmath}
\usepackage{graphicx}
\usepackage{amsfonts}
\usepackage{amssymb}
\usepackage{fullpage}%
\newtheorem{theorem}{Theorem}

\newtheorem{proposition}[theorem]{Proposition}

\newenvironment{proof}[1][Proof]{\textbf{#1.} }{\ \rule{0.5em}{0.5em}}
\begin{document}

\title{Quantum Computing and Hidden Variables I: Mapping Unitary to Stochastic Matrices}
\author{Scott Aaronson\thanks{University of California, Berkeley. \ Email:
aaronson@cs.berkeley.edu. \ Supported by an NSF Graduate Fellowship and by
DARPA grant F30602-01-2-0524.}}
\date{}
\maketitle

\begin{abstract}
This paper initiates the study of hidden variables from the discrete, abstract
perspective of quantum computing. \ For us, a hidden-variable theory is simply
a way to convert a unitary matrix that maps one quantum state to another, into
a stochastic matrix that maps the initial probability distribution to the
final one in some fixed basis. \ We list seven axioms that we might want such
a theory to satisfy, and then investigate which of the axioms can be satisfied
simultaneously. \ Toward this end, we construct a new hidden-variable theory
that is both robust to small perturbations and indifferent to the identity
operation, by exploiting an unexpected connection between unitary matrices and
network flows. \ We also analyze previous hidden-variable theories of Dieks
and Schr\"{o}dinger in terms of our axioms. \ In a companion paper, we will
show that actually \textit{sampling} the history of a hidden variable under
reasonable axioms is at least as hard as solving the Graph Isomorphism
problem; and indeed is probably intractable even for quantum computers.

\end{abstract}

\section{Introduction\label{INTRO}}

Quantum mechanics lets us calculate the probability that (say) an electron
will be found in an excited state if measured at a particular time. \ But it
is silent about \textit{multiple-time} or \textit{transition} probabilities:
that is, what is the probability that the electron will be in an excited state
at time $t_{1}$, given that it was in its ground state at an earlier time
$t_{0}$?\ \ The usual response is that this question is meaningless, unless of
course the electron was \textit{measured} (or otherwise known with probability
$1$) to be in its ground state at $t_{0}$. \ A different response---pursued by
Schr\"{o}dinger \cite{schrodinger}, Bohm \cite{bohm}, Bell \cite{bell}, Nelson
\cite{nelson}, Dieks \cite{dieks}, and others---treats the question as
provisionally meaningful, and then investigates how one might answer it
mathematically. \ Specific attempts at answers are called \textquotedblleft
hidden-variable theories.\textquotedblright

The appeal of hidden-variable theories is that they provide one possible
solution to the measurement problem. \ For they allow us to apply unitary
quantum mechanics to the entire universe (including ourselves), yet still
discuss the probability of a future observation conditioned on our current
observations. \ Furthermore, they let us do so without making any assumptions
about decoherence or the nature of observers. \ For example, even if an
observer were placed in coherent superposition, that observer would still have
a sequence of definite experiences, and the probability of any such sequence
could be calculated.

This paper initiates the study of hidden variables from a quantum computing
perspective. \ We restrict our attention to the simplest possible setting:
that of discrete time, a finite-dimensional Hilbert space, and a fixed
orthogonal basis. \ Within this setting, we reformulate known hidden-variable
theories due to Dieks \cite{dieks} and Schr\"{o}dinger \cite{schrodinger}, and
also introduce a new theory based on network flows. \ However, our main
contribution is the \textit{axiomatic approach} that we use. \ We propose
seven axioms for hidden-variable theories in our setting, and then compare
theories against each other based on which of the axioms they satisfy. \ A
central question in this approach is which subsets of axioms can be satisfied simultaneously.

In a companion paper \cite{aaronson}, we will make the connection to quantum
computing explicit, by proving that\ under any hidden-variable theory that
satisfies three reasonable axioms (called symmetry, indifference, and
robustness), the ability to examine one's entire \textquotedblleft
history\textquotedblright\ through a quantum system would entail the ability
to solve the Graph Isomorphism problem in polynomial time\textit{.} \ What
makes this result surprising is that, in the so-called oracle or black-box
model, sampling histories would \textit{not} entail the ability to solve
$\mathsf{NP}$-complete problems in polynomial time. \ We thus obtain the first
good example of a computational model that appears \textquotedblleft
slightly\textquotedblright\ more powerful than the usual quantum computing model.

This paper lays the groundwork for the computational results of
\cite{aaronson}, in particular by showing that there \textit{exists} a
hidden-variable theory satisfying the symmetry, indifference, and robustness axioms.

The paper is organized as follows. \ Section \ref{HV} formally defines
hidden-variable theories in our sense; then Section \ref{BOHM} contrasts these
theories with related ideas such as Bohmian mechanics and modal
interpretations. \ Section \ref{OBJECTIONS} addresses the most common
objections to\ our approach: for example, that the implicit dependence on a
fixed basis is unacceptable.

In Section \ref{AXIOMS}, we introduce seven possible axioms for
hidden-variable theories. \ These are symmetry under permutation of basis
states; indifference to the identity operation; robustness to small
perturbations; \textquotedblleft block robustness,\textquotedblright\ a weaker
version of robustness; commutativity with respect to spacelike-separated
unitaries; commutativity for the special case of product states; and
invariance under decomposition of mixed states into pure states. \ Ideally, a
theory would satisfy all of these axioms. \ However, we show in Section
\ref{IMPOS}\ that no theory satisfies both indifference and commutativity; no
theory satisfies both indifference and a stronger version of robustness; no
theory satisfies indifference, robustness, and decomposition invariance; and
no theory satisfies a stronger version of decomposition invariance. \ The
proofs of two of these results use the same geometric facts that underlie the
Bell inequalities.

In Section \ref{SPECIFIC}\ we shift from negative to positive results.
\ Section \ref{FLOW}\ presents a hidden-variable theory called the
\textit{flow theory} or $\mathcal{FT}$, which is based on the Max-Flow-Min-Cut
theorem from combinatorial optimization. \ The idea is to define a network of
\textquotedblleft pipes\textquotedblright\ from basis states at an initial
time to basis states at a final time, and then route as much probability mass
as possible through these pipes. \ The capacity of each pipe depends on the
corresponding entry of the unitary acting from the initial to final time. \ To
find the probability of transitioning from basis state $\left\vert
i\right\rangle $ to basis state $\left\vert j\right\rangle $, we then
determine how much of the flow originating at $\left\vert i\right\rangle $\ is
routed along the pipe to $\left\vert j\right\rangle $. \ Our main results are
that $\mathcal{FT}$ is well-defined and that it is robust to small
perturbations. \ Since $\mathcal{FT}$ trivially satisfies the indifference
axiom, this implies that the indifference and robustness axioms can be
satisfied simultaneously, which was not at all obvious\ \textit{a priori}.
\ The flow theory also satisfies symmetry, but not product commutativity,
scalar invariance, or decomposition invariance.

Section \ref{SCHROD}\ presents a second theory that we call the
\textit{Schr\"{o}dinger theory} or $\mathcal{ST}$, since it is based on a pair
of integral equations introduced in a 1931 paper of Schr\"{o}dinger
\cite{schrodinger}. \ Schr\"{o}dinger conjectured, but was unable to prove,
the existence and uniqueness of a solution to these equations; the problem was
not settled until the work of Nagasawa \cite{nagasawa}\ in the 1980's. \ In
our discrete setting the problem is simpler, and we give a self-contained
proof of existence using a matrix scaling technique due to Sinkhorn
\cite{sinkhorn}. \ The idea is as follows: we want to convert a unitary matrix
that maps one quantum state to another, into a nonnegative matrix whose
$i^{th}$\ column sums to the initial probability of basis state $\left\vert
i\right\rangle $, and whose $j^{th}$\ row sums to the final probability of
basis state $\left\vert j\right\rangle $. \ To do so, we first replace each
entry of the unitary matrix by its absolute value, then normalize each column
to sum to the desired initial probability, then normalize each row to sum to
the desired final probability. \ But then the columns are no longer normalized
correctly, so we normalize them \textit{again}, then normalize the rows again,
and so on. \ We show that this iterative process converges, from which it
follows that $\mathcal{ST}$ is well-defined. \ We also show that
$\mathcal{ST}$\ satisfies the symmetry, indifference, and product
commutativity axioms; and violates the decomposition invariance axiom. \ We
conjecture that $\mathcal{ST}$ satisfies the robustness axiom; proving that
conjecture is the main open problem of the paper. \ We conclude in Section
\ref{DISC}.

\subsection{Hidden-Variable Theories\label{HV}}

Suppose we have an $N\times N$\ unitary matrix $U$, acting on a state%
\[
\left\vert \psi\right\rangle =\alpha_{1}\left\vert 1\right\rangle
+\cdots+\alpha_{N}\left\vert N\right\rangle ,
\]
where $\left\vert 1\right\rangle ,\ldots,\left\vert N\right\rangle $\ is a
standard orthogonal basis. \ Let%
\[
U\left\vert \psi\right\rangle =\beta_{1}\left\vert 1\right\rangle
+\cdots+\beta_{N}\left\vert N\right\rangle .
\]
Then can we construct a stochastic matrix $S$, which maps the vector of
probabilities%
\[
\overrightarrow{p}=\left[
\begin{array}
[c]{c}%
\left\vert \alpha_{1}\right\vert ^{2}\\
\vdots\\
\left\vert \alpha_{N}\right\vert ^{2}%
\end{array}
\right]
\]
induced by measuring $\left\vert \psi\right\rangle $, to the vector%
\[
\overrightarrow{q}=\left[
\begin{array}
[c]{c}%
\left\vert \beta_{1}\right\vert ^{2}\\
\vdots\\
\left\vert \beta_{N}\right\vert ^{2}%
\end{array}
\right]
\]
induced by measuring $U\left\vert \psi\right\rangle $? \ Trivially yes. \ The
following matrix maps \textit{any} vector of probabilities to $\overrightarrow
{q}$, ignoring the input vector $\overrightarrow{p}$\ entirely:
\[
S_{\mathcal{PT}}=\left[
\begin{array}
[c]{ccc}%
\left\vert \beta_{1}\right\vert ^{2} & \cdots & \left\vert \beta
_{1}\right\vert ^{2}\\
\vdots &  & \vdots\\
\left\vert \beta_{N}\right\vert ^{2} & \cdots & \left\vert \beta
_{N}\right\vert ^{2}%
\end{array}
\right]  .
\]
Here $\mathcal{PT}$\ stands for \textit{product theory}. \ The product theory
corresponds to a strange picture of physical reality, in which memories and
records are completely unreliable, there being no causal connection between
states of affairs at earlier and later times.

So we would like $S$ to depend on $U$\ itself somehow, not just on $\left\vert
\psi\right\rangle $\ and $U\left\vert \psi\right\rangle $. \ Indeed, ideally
$S$\ would be a function \textit{only} of $U$, and not of $\left\vert
\psi\right\rangle $. \ But this is impossible, as the following example shows.
\ Let $U$ be a $\pi/4$\ rotation, and let $\left\vert +\right\rangle =\left(
\left\vert 0\right\rangle +\left\vert 1\right\rangle \right)  /\sqrt{2}$ and
$\left\vert -\right\rangle =\left(  \left\vert 0\right\rangle -\left\vert
1\right\rangle \right)  /\sqrt{2}$. \ Then $U\left\vert +\right\rangle
=\left\vert 1\right\rangle $\ implies that%
\[
S\left(  \left\vert +\right\rangle ,U\right)  =\left[
\begin{array}
[c]{cc}%
0 & 0\\
1 & 1
\end{array}
\right]  ,
\]
whereas $U\left\vert -\right\rangle =\left\vert 0\right\rangle $\ implies that%
\[
S\left(  \left\vert -\right\rangle ,U\right)  =\left[
\begin{array}
[c]{cc}%
1 & 1\\
0 & 0
\end{array}
\right]  .
\]

On the other hand, it is easy to see that, if $S$ can depend on $\left\vert
\psi\right\rangle $ as well as $U$, then there are infinitely many choices for
the function $S\left(  \left\vert \psi\right\rangle ,U\right)  $. \ Every
choice reproduces the predictions of quantum mechanics perfectly when
restricted to single-time probabilities. \ So how can we possibly choose among
them? \ Our approach in Sections \ref{AXIOMS}\ and \ref{SPECIFIC} will be to
write down axioms that we would like $S$ to satisfy, and then investigate
which of the axioms can be satisfied simultaneously.

Formally, a \textit{hidden-variable theory} is a family of functions $\left\{
S_{N}\right\}  _{N\geq1}$, where each $S_{N}$\ maps an $N$-dimensional mixed
state $\rho$\ and an $N\times N$\ \ unitary matrix $U$\ onto a singly
stochastic matrix $S_{N}\left(  \rho,U\right)  $. \ We will often suppress the
dependence on $N$, and use subscripts such as $\mathcal{PT}$\ or
$\mathcal{FT}$\ to indicate the theory in question. \ Also, if\ $\rho
=\left\vert \psi\right\rangle \left\langle \psi\right\vert $\ is a pure state
we may write $S\left(  \left\vert \psi\right\rangle ,U\right)  $\ instead of
$S\left(  \left\vert \psi\right\rangle \left\langle \psi\right\vert ,U\right)
$.

Let $\left(  M\right)  _{ij}$\ denote the entry in the $i^{th}$\ column and
$j^{th}$\ row of matrix $M$. \ Then $\left(  S\right)  _{ij}$ is the
probability that the hidden variable takes value $\left\vert j\right\rangle $
after $U$\ is applied, conditioned on it taking value $\left\vert
i\right\rangle $\ before $U$ is applied. \ At a minimum, any theory must
satisfy the following marginalization axiom: for all $j\in\left\{
1,\ldots,N\right\}  $,\vspace{0pt}%
\[
\sum_{i}\left(  S\right)  _{ij}\left(  \rho\right)  _{ij}=\left(  U\rho
U^{-1}\right)  _{jj}\text{.}%
\]
This says that after $U$\ is applied, the hidden variable takes value
$\left\vert j\right\rangle $\ with probability $\left(  U\rho U^{-1}\right)
_{jj}$, which is the usual Born probability.

Often it will be convenient to refer, not to $S$ itself, but to the matrix
$P\left(  \rho,U\right)  $ of joint probabilities\ whose $\left(  i,j\right)
$\ entry is $\left(  P\right)  _{ij}=\left(  S\right)  _{ij}\left(
\rho\right)  _{ii}$. \ The $i^{th}$\ column of $P$\ must sum to $\left(
\rho\right)  _{ii}$, and the $j^{th}$\ row must sum to $\left(  U\rho
U^{-1}\right)  _{jj}$. \ Indeed, we will define the theories $\mathcal{FT}%
$\ and $\mathcal{ST}$\ by first specifying the matrix $P$, and then setting
$\left(  S\right)  _{ij}:=\left(  P\right)  _{ij}/\left(  \rho\right)  _{ii}$.
\ This approach has the drawback that if $\left(  \rho\right)  _{ii}=0$, then
the $i^{th}$\ column of $S$ is undefined. \ To get around this, we adopt the
convention that%
\[
S\left(  \rho,U\right)  :=\lim_{\varepsilon\rightarrow0^{+}}S\left(
\rho_{\varepsilon},U\right)
\]
where $\rho_{\varepsilon}=\left(  1-\varepsilon\right)  \rho+\varepsilon
I$\ and $I$ is the $N\times N$\ maximally mixed state. \ Technically, the
limits%
\[
\lim_{\varepsilon\rightarrow0^{+}}\frac{\left(  P\left(  \rho_{\varepsilon
},U\right)  \right)  _{ij}}{\left(  \rho_{\varepsilon}\right)  _{ii}}%
\]
might not exist, but in the cases of interest to us it will be obvious that
they do.

\subsection{Comparison with Previous Work\label{BOHM}}

Before going further, we should contrast our approach with previous approaches
to hidden variables, the most famous of which is Bohmian mechanics
\cite{bohm}. \ Our main criticism of Bohmian mechanics is that it commits
itself to a Hilbert space of particle positions and momenta. \ Furthermore, it
is crucial that the positions and momenta be \textit{continuous}, in order for
particles to evolve deterministically. \ To see this, let $\left\vert
L\right\rangle $\ and $\left\vert R\right\rangle $\ be discrete positions, and
suppose a particle is in state $\left\vert L\right\rangle $\ at time $t_{1}$,
and state $\left(  \left\vert L\right\rangle +\left\vert R\right\rangle
\right)  /\sqrt{2}$ at a later time $t_{2}$. \ Then a hidden variable
representing the position would have entropy $0$ at $t_{1}$, since it is
always $\left\vert L\right\rangle $ then; but entropy $1$\ at $t_{2}$, since
it is $\left\vert L\right\rangle $\ or $\left\vert R\right\rangle $\ both with
$1/2$ probability. \ Therefore the earlier value cannot determine the later
one.\footnote{Put differently, Bohm's conservation of probability result
breaks down because the \textquotedblleft wavefunctions\textquotedblright\ at
$t_{1}$\ and $t_{2}$ are degenerate, with all amplitude concentrated on
finitely many points. \ But in a discrete Hilbert space, \textit{every}
wavefunction is degenerate in this sense!} \ It follows that Bohmian mechanics
is incompatible with the belief that\ all physical observables are discrete.
\ But in our view, there are strong reasons to hold that belief, which include
black hole entropy bounds; the existence of a natural minimum length scale
($10^{-33}$ cm); results on area quantization in quantum gravity \cite{rs};
the fact that many physical quantities once thought to be continuous have
turned out to be discrete; the infinities of quantum field theory;\ the
implausibility of analog \textquotedblleft hypercomputers\textquotedblright;
and conceptual problems raised by the independence of the continuum hypothesis.

Of course there exist stochastic analogues of Bohmian mechanics, among them
Nelsonian mechanics \cite{nelson}\ and Bohm and Hiley's \textquotedblleft
stochastic interpretation\textquotedblright\ \cite{bh}. \ But it is not
obvious why we should prefer these to other stochastic hidden-variable
theories. \ From a quantum-information perspective, it is much more natural to
take an abstract approach---one that allows arbitrary finite-dimensional
Hilbert spaces, and that does not rule out any transition rule \textit{a
priori}.

Stochastic hidden variables have also been considered in the context of modal
interpretations; see Dickson \cite{dickson}, Bacciagaluppi and Dickson
\cite{bd},\ and Dieks \cite{dieks}\ for example. \ However, the central
assumptions in that work are extremely different from ours. \ In modal
interpretations, a pure state evolving unitarily poses no problems at all: one
simply rotates the hidden-variable basis along with the state, so that the
state always represents a \textquotedblleft possessed
property\textquotedblright\ of the system in the current basis. \ Difficulties
arise only for mixed states; and there, the goal is to track a whole set of
possessed properties. \ By contrast, our approach is to fix an orthogonal
basis, then track a single hidden variable that is an element of that basis.
\ The issues raised by pure states and mixed states are essentially the same.

Finally we should mention the consistent-histories interpretation of Griffiths
\cite{griffiths} and Gell-Mann and Hartle \cite{gh}. \ This interpretation
assigns probabilities to various histories through a quantum system, so long
as the \textquotedblleft interference\textquotedblright\ between those
histories is negligible. \ Loosely speaking, then, the situations where
consistent histories make sense are precisely the ones where the question of
transition probabilities can be avoided.

\subsection{Objections\label{OBJECTIONS}}

Hidden-variable theories, as we define them, are open to several technical
objections. \ For example, we required transition probabilities for only one
orthogonal observable. \ What about other observables? \ The problem is that,
according to the Kochen-Specker theorem,\ we cannot assign consistent values
to all observables at any \textit{single} time, let alone give transition
probabilities for those values. \ This is an issue in any setting, not just
ours. \ The solution we prefer is to postulate a fixed orthogonal basis of
\textquotedblleft distinguishable experiences,\textquotedblright\ and to
interpret a measurement in any other basis as a unitary followed by a
measurement in the fixed basis. \ As mentioned in Section \ref{BOHM}, modal
interpretations opt for a different solution, which involves sets of bases
that change over time with the state itself. \ It might be interesting to
combine the approaches.

Another objection is that the probability of transitioning from basis state
$\left\vert i\right\rangle $\ at time $t_{1}$\ to basis state $\left\vert
j\right\rangle $\ at time $t_{2}$\ might depend on how finely\ we divide the
time interval between $t_{1}$\ and $t_{2}$. \ In other words, for some state
$\left\vert \psi\right\rangle $\ and unitaries $V,W$, we might have%
\[
S\left(  \left\vert \psi\right\rangle ,WV\right)  \neq S\left(  V\left\vert
\psi\right\rangle ,W\right)  S\left(  \left\vert \psi\right\rangle ,V\right)
\]
(a similar point was made by Gillespie \cite{gillespie}). \ Indeed, this is
true for any hidden-variable theory other than the product theory
$\mathcal{PT}$. \ To see this, observe that for all unitaries $U$ and states
$\left\vert \psi\right\rangle $, there exist unitaries $V,W$\ such that
$U=WV$\ and $V\left\vert \psi\right\rangle =\left\vert 1\right\rangle $.
\ Then applying $V$\ destroys all information in the hidden variable (that is,
decreases its entropy to $0$); so if we then apply $W$, then the variable's
final value must be uncorrelated with the initial value. \ In other words,
$S\left(  V\left\vert \psi\right\rangle ,W\right)  S\left(  \left\vert
\psi\right\rangle ,V\right)  $\ must equal $S_{\mathcal{PT}}\left(  \left\vert
\psi\right\rangle ,U\right)  $.\ \ It follows that to any hidden-variable
theory we must associate a time scale, or some other rule for deciding when
the transitions take place.

In our defense, let us point out that exactly the same problem arises in
\textit{continuous}-time stochastic hidden-variable theories. \ For if a state
$\left\vert \psi\right\rangle $\ is governed by the Schr\"{o}dinger equation
$d\left\vert \psi\right\rangle /dt=iH_{t}\left\vert \psi\right\rangle $, and a
hidden variable's probability distribution $\overrightarrow{p}$ is governed by
the stochastic equation $d\overrightarrow{p}/d\tau=A_{\tau}\overrightarrow{p}%
$, then there is still an arbitrary parameter $d\tau/dt$ on which the dynamics depend.

Finally, it will be objected that we have ignored special relativity. \ In
Section \ref{AXIOMS}\ we will define a \textit{commutativity axiom}, which
informally requires that the stochastic matrix $S$ not depend on the temporal
order of spacelike separated events. \ Unfortunately, we will see that when
entangled states are involved, commutativity is irreconcilable with another
axiom that seems even more basic. \ The resulting nonlocality has the same
character as the nonlocality of Bohmian mechanics---that is, one cannot use it
to send superluminal signals in the usual sense, but it is unsettling nonetheless.

\section{Axioms for Hidden-Variable Theories\label{AXIOMS}}

We now state seven axioms that we would like hidden-variable theories to satisfy.

\textbf{Symmetry.} \ A theory is symmetric if it is invariant under relabeling
of basis states: that is, if for all $\rho,U$\ and all permutation matrices
$Q$,\vspace{0pt}%
\[
Q^{-1}S\left(  \rho,U\right)  Q=S\vspace{0pt}\left(  Q^{-1}\rho Q,Q^{-1}%
UQ\right)  \text{.}%
\]
All theories discussed in this paper are symmetric.

\textbf{Indifference.} \ Suppose we partition the basis states\ into `blocks,'
between which $U$ can never produce interference. \ Call an ordered pair
$\left\langle I,J\right\rangle $ of subsets of $\left\{  1,\ldots,N\right\}  $
a \textit{block} if $\left(  U\right)  _{ij}=0$ for all $i\in I$\ and $j\notin
J$, as well as all $i\notin I$\ and $j\in J$. \ Also, call $\left\langle
I,J\right\rangle $\ a \textit{minimal block} if $\left\vert I\right\vert
=\left\vert J\right\vert $\ and no $\left\langle I^{\ast},J^{\ast
}\right\rangle $\ with $I^{\ast}\subset I$\ and $J^{\ast}\subset J$ is a
block. \ Note that if $\left\langle I_{1},J_{1}\right\rangle ,\ldots
,\left\langle I_{m},J_{m}\right\rangle $\ are the minimal blocks, then both
$\left\{  I_{1},\ldots,I_{m}\right\}  $ and $\left\{  J_{1},\ldots
,J_{m}\right\}  $ are partitions of $\left\{  1,\ldots,N\right\}  $. \ We say
a theory is indifferent if it never produces interference between minimal
blocks---that is, if $\left(  S\right)  _{ij}=0$\ for all $i,j$\ in different
minimal blocks. \ In particular, indifference implies that given any state
$\rho$\ in a tensor product space $\mathcal{H}_{A}\otimes\mathcal{H}_{B}$, and
any unitary $U$ that acts only on $\mathcal{H}_{A}$\ (that is, is the identity
on $\mathcal{H}_{B}$), the stochastic matrix $S\left(  \rho,U\right)  $\ acts
only on $\mathcal{H}_{A}$\ as well.

\textbf{Robustness.} \ A theory is robust if it is insensitive to small errors
in a state or unitary (which, in particular, implies continuity). \ Suppose we
obtain $\widetilde{\rho}$ and $\widetilde{U}$\ by making small changes to
$\rho$\ and $U$\ respectively.\ \ Then for all polynomials $p$, there should
exist a polynomial $q$\ such that for all $N$,%
\[
\vspace{0pt}\left\Vert P\left(  \widetilde{\rho},\widetilde{U}\right)
-P\left(  \rho,U\right)  \right\Vert \leq\frac{1}{p\left(  N\right)  }%
\]
where $\left\Vert M\right\Vert =\max_{ij}\left\vert \left(  M\right)
_{ij}\right\vert $, whenever $\left\Vert \widetilde{\rho}-\rho\right\Vert
\leq1/q\left(  N\right)  $ and $\left\Vert \widetilde{U}-U\right\Vert
\leq1/q\left(  N\right)  $.\ \ Robustness has an important advantage for
quantum computing: if a hidden-variable theory is robust then the set of gates
used to define the unitaries $U_{1},\ldots,U_{T}$\ is irrelevant, since by the
Solovay-Kitaev Theorem (see \cite{nc}), any universal quantum gate set can
simulate any other to a precision $\varepsilon$\ with $O\left(  \log
^{c}1/\varepsilon\right)  $\ overhead.

\textbf{Block Robustness.} \ Unfortunately, one of the theories that we wish
to study does not satisfy robustness. \ We therefore define a weaker notion of
robustness that this theory satisfies. \ We say a hidden-variable theory is
\textit{block robust} if robustness holds for all modifications $\rho
,U\rightarrow\widetilde{\rho},\widetilde{U}$\ such that $U$ and $\widetilde
{U}$\ have the same set $\left\langle I_{1},J_{1}\right\rangle ,\ldots
,\left\langle I_{m},J_{m}\right\rangle $\ of minimal blocks.

\textbf{Commutativity.} \ Let $\rho_{AB}$\ be a bipartite state, and let
$U_{A}$ and $U_{B}$\ act only on subsystems $A$ and $B$ respectively. \ Then
commutativity means that the order in which $U_{A}$ and $U_{B}$ are applied is
irrelevant:%
\[
S\left(  U_{A}\rho_{AB}U_{A}^{-1},U_{B}\right)  S\left(  \rho_{AB}%
,U_{A}\right)  \vspace{0pt}=S\left(  U_{B}\rho_{AB}U_{B}^{-1},U_{A}\right)
S\left(  \rho_{AB},U_{B}\right)  \text{.}%
\]

\textbf{Product Commutativity.} \ A theory is product commutative if it
satisfies commutativity for all separable pure states $\left\vert
\psi\right\rangle =\left\vert \psi_{A}\right\rangle \otimes\left\vert \psi
_{B}\right\rangle $.

\textbf{Decomposition Invariance.} \ A theory is decomposition invariant if%

\[
S\left(  \rho,U\right)  =\sum_{i=1}^{N}p_{i}S\left(  \left\vert \psi
_{i}\right\rangle \left\langle \psi_{i}\right\vert ,U\right)
\]
for every decomposition%
\[
\rho=\sum_{i=1}^{N}p_{i}\left\vert \psi_{i}\right\rangle \left\langle \psi
_{i}\right\vert
\]
of $\rho$\ into pure states. \ Theorem \ref{decomp}, part (ii) will show that
the analogous axiom for $P\left(  \rho,U\right)  $\ is unsatisfiable.

\subsection{Comparing Hidden-Variable Theories\label{COMPARE}}

To fix ideas, let us compare some hidden-variable theories with respect to the
above axioms. \ We have already seen the product theory $\mathcal{PT}$\ in
Section \ref{HV}. \ It is easy to show that $\mathcal{PT}$\ satisfies
symmetry, robustness, commutativity, and decomposition invariance. \ However,
we consider $\mathcal{PT}$ unsatisfactory because it violates indifference:
even if a unitary $U$ acts only on the first of two qubits, $S_{\mathcal{PT}%
}\left(  \rho,U\right)  $\ will readily produce transitions between (say)
$\left\vert 00\right\rangle $\ and $\left\vert 01\right\rangle $, or between
$\left\vert 01\right\rangle $\ and $\left\vert 10\right\rangle $.

Recognizing this problem, Dieks \cite{dieks}\ proposed an alternative theory
that in our terminology corresponds to the following.\footnote{Dieks (personal
communication) says he would no longer defend this theory.} \ First find the
minimal blocks $\left\langle I_{1},J_{1}\right\rangle ,\ldots,\left\langle
I_{m},J_{m}\right\rangle $ of $U$. \ Then apply the product theory separately
to each minimal block; that is, if $i$ and $j$ belong to the same block
$\left\langle I,J\right\rangle $\ then set%
\[
\left(  S\right)  _{ij}=\frac{\left(  U\rho U^{-1}\right)  _{jj}}%
{\sum_{\widehat{j}\in J}\left(  U\rho U^{-1}\right)  _{\widehat{j}\widehat{j}%
}},
\]
and otherwise set $\left(  S\right)  _{ij}=0$. \ The resulting \textit{Dieks
theory}, $\mathcal{DT}$,\ clearly satisfies indifference. \ However, it does
not satisfy robustness (or even continuity), since the set of minimal blocks
can change if we replace `$0$' entries in $U$ by arbitrarily small nonzero entries.

In Section \ref{SPECIFIC}\ we will introduce two other hidden-variable
theories, the flow theory $\mathcal{FT}$\ and the Schr\"{o}dinger theory
$\mathcal{ST}$. \ The following table lists which axioms the four theories satisfy.%

\begin{tabular}
[c]{lllll}
& $\mathcal{PT}$ (Product) & $\mathcal{DT}$ (Dieks) & $\mathcal{FT}$ (Flow) &
$\mathcal{ST}$ (Schr\"{o}dinger)\\
Symmetry & Yes & Yes & Yes & Yes\\
Indifference & No & Yes & Yes & Yes\\
Robustness & Yes & No & Yes & ?\\
Block Robustness & Yes & Yes & Yes & ?\\
Commutativity & Yes & No & No & No\\
Product Commutativity & Yes & Yes & No & Yes\\
Decomposition Invariance & Yes & Yes & No & No
\end{tabular}

If we could prove that $\mathcal{ST}$\ satisfies robustness, then the above
table together with the impossibility results of Section \ref{IMPOS}\ would
completely characterize which of the axioms can be satisfied simultaneously.

\section{Impossibility Results\label{IMPOS}}

This section shows that certain sets of axioms cannot be satisfied by any
hidden-variable theory. \ We first show that the failure of $\mathcal{DT}%
$,\ $\mathcal{FT}$, and $\mathcal{ST}$\ to satisfy commutativity is inherent,
and not a fixable technical problem.

\begin{theorem}
\label{nogo}No hidden-variable theory satisfies both indifference and commutativity.
\end{theorem}

\begin{proof}
Assume indifference holds, and let our initial state be $\left\vert
\psi\right\rangle =\left(  \left\vert 00\right\rangle +\left\vert
11\right\rangle \right)  /\sqrt{2}$. \ Suppose $U_{A}$\ applies a $\pi/8$
rotation to the first qubit, and $U_{B}$\ applies a\ $-\pi/8$\ rotation to the
second qubit. \ Then%
\begin{align*}
U_{A}\left\vert \psi\right\rangle  &  =U_{B}\left\vert \psi\right\rangle
=\frac{1}{\sqrt{2}}\left(  \cos\frac{\pi}{8}\left\vert 00\right\rangle
-\sin\frac{\pi}{8}\left\vert 01\right\rangle +\sin\frac{\pi}{8}\left\vert
10\right\rangle +\cos\frac{\pi}{8}\left\vert 11\right\rangle \right)  ,\\
U_{A}U_{B}\left\vert \psi\right\rangle  &  =U_{B}U_{A}\left\vert
\psi\right\rangle =\frac{1}{2}\left(  \left\vert 00\right\rangle -\left\vert
01\right\rangle +\left\vert 10\right\rangle +\left\vert 11\right\rangle
\right)  .
\end{align*}
Let $v_{t}$\ be the value of the hidden variable after $t$ unitaries have been
applied. \ Let $E$ be the event that $v_{0}=\left\vert 00\right\rangle
$\ initially, and $v_{2}=\left\vert 10\right\rangle $\ at the end.\ \ If
$U_{A}$\ is applied before $U_{B}$,\ then the unique `path' from $v_{0}$\ to
$v_{2}$\ consistent with indifference sets $v_{1}=\left\vert 10\right\rangle
$. \ So%
\[
\Pr\left[  E\right]  \leq\Pr\left[  v_{1}=\left\vert 10\right\rangle \right]
=\frac{1}{2}\sin^{2}\frac{\pi}{8}.
\]
But if $U_{B}$\ is applied before $U_{A}$, then the probability that
$v_{0}=\left\vert 11\right\rangle $\ and $v_{2}=\left\vert 10\right\rangle
$\ is at most $\frac{1}{2}\sin^{2}\frac{\pi}{8}$, by the same reasoning.
\ Thus, since $v_{2}$\ must equal $\left\vert 10\right\rangle $\ with
probability $1/4$, and since the only possibilities for $v_{0}$\ are
$\left\vert 00\right\rangle $\ and $\left\vert 11\right\rangle $,%
\[
\Pr\left[  E\right]  \geq\frac{1}{4}-\frac{1}{2}\sin^{2}\frac{\pi}{8}>\frac
{1}{2}\sin^{2}\frac{\pi}{8}.
\]
We conclude that commutativity is violated.
\end{proof}

Let us remark on the relationship between Theorem \ref{nogo}\ and Bell's
Theorem. \ Any hidden-variable theory that is \textquotedblleft
local\textquotedblright\ in Bell's sense would immediately satisfy both
indifference and commutativity. \ However, the converse is not obvious, since
there might be nonlocal information in the states $U_{A}\left\vert
\psi\right\rangle $ or $U_{B}\left\vert \psi\right\rangle $,\ which an
indifferent commutative theory could exploit but a local one could not.
\ Theorem \ref{nogo} rules out this possibility, and in that sense is a
strengthening of Bell's Theorem.

The next result places limits on decomposition invariance.

\begin{theorem}
\label{decomp}\quad

\begin{enumerate}
\item[(i)] No theory satisfies indifference, robustness, and decomposition invariance.

\item[(ii)] No theory has the property that%
\[
P\left(  \rho,U\right)  =\sum_{i=1}^{N}p_{i}P\left(  \left\vert \psi
_{i}\right\rangle \left\langle \psi_{i}\right\vert ,U\right)
\]
for every decomposition $\sum_{i=1}^{N}p_{i}\left\vert \psi_{i}\right\rangle
\left\langle \psi_{i}\right\vert $\ of $\rho$.
\end{enumerate}
\end{theorem}

\begin{proof}
\quad

\begin{enumerate}
\item[(i)] Suppose the contrary. \ Let%
\begin{align*}
R_{\theta}  &  =\left[
\begin{array}
[c]{cc}%
\cos\theta & -\sin\theta\\
\sin\theta & \cos\theta
\end{array}
\right]  ,\\
\left\vert \varphi_{\theta}\right\rangle  &  =\cos\theta\left\vert
0\right\rangle +\sin\theta\left\vert 1\right\rangle .
\end{align*}
Then for every $\theta$ not a multiple of $\pi/2$, we must have%
\begin{align*}
S\left(  \left\vert \varphi_{-\theta}\right\rangle ,R_{\theta}\right)   &
=\left[
\begin{array}
[c]{cc}%
1 & 1\\
0 & 0
\end{array}
\right]  ,\\
S\left(  \left\vert \varphi_{\pi/2-\theta}\right\rangle ,R_{\theta}\right)
&  =\left[
\begin{array}
[c]{cc}%
0 & 0\\
1 & 1
\end{array}
\right]  .
\end{align*}
So by decomposition invariance, letting $I=\left(  \left\vert 0\right\rangle
\left\langle 0\right\vert +\left\vert 1\right\rangle \left\langle 1\right\vert
\right)  /2$\ denote the maximally mixed state,%
\[
S\left(  I,R_{\theta}\right)  =S\left(  \frac{\left\vert \varphi_{-\theta
}\right\rangle \left\langle \varphi_{-\theta}\right\vert +\left\vert
\varphi_{\pi/2-\theta}\right\rangle \left\langle \varphi_{\pi/2-\theta
}\right\vert }{2},R_{\theta}\right)  =\left[
\begin{array}
[c]{cc}%
\frac{1}{2} & \frac{1}{2}\\
\frac{1}{2} & \frac{1}{2}%
\end{array}
\right]
\]
and therefore%
\[
P\left(  I,R_{\theta}\right)  =\left[
\begin{array}
[c]{cc}%
\frac{\left(  \rho\right)  _{00}}{2} & \frac{\left(  \rho\right)  _{11}}{2}\\
\frac{\left(  \rho\right)  _{00}}{2} & \frac{\left(  \rho\right)  _{11}}{2}%
\end{array}
\right]  =\left[
\begin{array}
[c]{cc}%
\frac{1}{4} & \frac{1}{4}\\
\frac{1}{4} & \frac{1}{4}%
\end{array}
\right]  .
\]
By robustness, this holds for $\theta=0$ as well. \ \ But this is a
contradiction, since by indifference $P\left(  I,R_{0}\right)  $\ must be half
the identity.

\item[(ii)] Reminiscent of Theorem \ref{nogo}. \ Suppose the contrary; then%
\[
P\left(  I,R_{\pi/8}\right)  =\frac{P\left(  \left\vert 0\right\rangle
,R_{\pi/8}\right)  +P\left(  \left\vert 1\right\rangle ,R_{\pi/8}\right)  }%
{2}.
\]
So considering transitions from $\left\vert 0\right\rangle $\ to $\left\vert
1\right\rangle $,%
\[
\left(  P\left(  I,R_{\pi/8}\right)  \right)  _{01}=\frac{\left(  P\left(
\left\vert 0\right\rangle ,R_{\pi/8}\right)  \right)  _{11}+0}{2}=\frac{1}%
{2}\sin^{2}\frac{\pi}{8}.
\]
But%
\[
P\left(  I,R_{\pi/8}\right)  =\frac{P\left(  \left\vert \varphi_{\pi
/8}\right\rangle ,R_{\pi/8}\right)  +P\left(  \left\vert \varphi_{5\pi
/8}\right\rangle ,R_{\pi/8}\right)  }{2}%
\]
also. \ Since $R_{\pi/8}\left\vert \varphi_{\pi/8}\right\rangle =\left\vert
\varphi_{\pi/4}\right\rangle $, we have%
\begin{align*}
\left(  P\left(  I,R_{\pi/8}\right)  \right)  _{01}  &  \geq\frac{1}{2}\left(
P\left(  \left\vert \varphi_{\pi/8}\right\rangle ,R_{\pi/8}\right)  \right)
_{01}\\
&  \geq\frac{1}{2}\left(  \frac{1}{2}-\left(  P\left(  \left\vert \varphi
_{\pi/8}\right\rangle ,R_{\pi/8}\right)  \right)  _{11}\right) \\
&  \geq\frac{1}{2}\left(  \frac{1}{2}-\sin^{2}\frac{\pi}{8}\right) \\
&  >\frac{1}{2}\sin^{2}\frac{\pi}{8}%
\end{align*}
which is a contradiction.
\end{enumerate}
\end{proof}

Notice that all three conditions in Theorem \ref{decomp}, part (i) were
essential---for $\mathcal{PT}$\ satisfies robustness and decomposition
invariance, $\mathcal{DT}$\ satisfies indifference and decomposition
invariance, and $\mathcal{FT}$\ satisfies indifference and robustness.

Our last impossibility result says that no hidden-variable theory satisfies
both indifference and \textquotedblleft strong continuity,\textquotedblright%
\ in the sense that for all $\varepsilon>0$ there exists $\delta>0$\ such that
$\left\Vert \widetilde{\rho}-\rho\right\Vert \leq\delta$ implies $\left\Vert
S\left(  \widetilde{\rho},U\right)  -S\left(  \rho,U\right)  \right\Vert
\leq\varepsilon$. \ To see this, let%
\begin{align*}
U &  =\left[
\begin{array}
[c]{ccc}%
1 & 0 & 0\\
0 & \frac{1}{\sqrt{2}} & -\frac{1}{\sqrt{2}}\\
0 & \frac{1}{\sqrt{2}} & \frac{1}{\sqrt{2}}%
\end{array}
\right]  ,\\
\rho &  =\sqrt{1-2\delta^{2}}\left\vert 0\right\rangle +\delta\left\vert
1\right\rangle +\delta\left\vert 2\right\rangle ,\\
\widetilde{\rho} &  =\sqrt{1-2\delta^{2}}\left\vert 0\right\rangle
+\delta\left\vert 1\right\rangle -\delta\left\vert 2\right\rangle .
\end{align*}
Then by indifference,%
\[
S\left(  \rho,U\right)  =\left[
\begin{array}
[c]{ccc}%
1 & 0 & 0\\
0 & 0 & 0\\
0 & 1 & 1
\end{array}
\right]  ,~~~~~~~~S\left(  \widetilde{\rho},U\right)  =\left[
\begin{array}
[c]{ccc}%
1 & 0 & 0\\
0 & 1 & 1\\
0 & 0 & 0
\end{array}
\right]  .
\]
This is the reason why we defined robustness in terms of the joint
probabilities matrix $P$ rather than the stochastic matrix $S$. \ On the other
hand, note that by giving up indifference, we \textit{can} satisfy strong
continuity, as is shown by $\mathcal{FT}$.

\section{Specific Theories\label{SPECIFIC}}

This section presents the main results of the paper, which concern two
nontrivial examples of hidden-variable theories: the flow theory in Section
\ref{FLOW}, and the Schr\"{o}dinger\ theory in Section \ref{SCHROD}.

\subsection{Flow Theory\label{FLOW}}

\begin{figure}
[ptb]
\begin{center}
\includegraphics[
trim=1.414995in 3.647223in 2.031921in 0.608775in,
height=1.7783in,
width=3.4803in
]%
{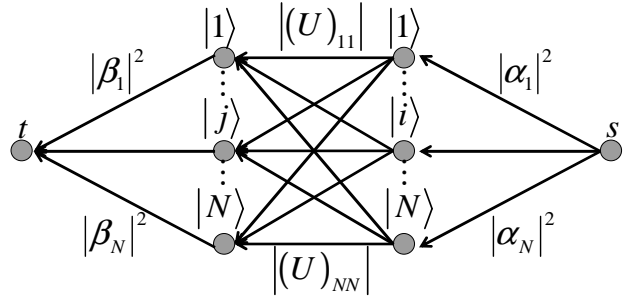}%
\caption{A network (weighted directed graph with source and sink)
corresponding to the unitary $U$ and state $\left\vert \psi\right\rangle $}%
\label{flowfig}%
\end{center}
\end{figure}

The idea of the flow theory is to convert a unitary matrix into a weighted
directed graph, and then route probability mass through that graph like oil
through pipes. \ Given a unitary $U$, let
\[
\left[
\begin{array}
[c]{c}%
\beta_{1}\\
\vdots\\
\beta_{N}%
\end{array}
\right]  =\left[
\begin{array}
[c]{ccc}%
\left(  U\right)  _{11} & \cdots & \left(  U\right)  _{N1}\\
\vdots &  & \vdots\\
\left(  U\right)  _{1N} & \cdots & \left(  U\right)  _{NN}%
\end{array}
\right]  \left[
\begin{array}
[c]{c}%
\alpha_{1}\\
\vdots\\
\alpha_{N}%
\end{array}
\right]  ,
\]
where for the time being%
\begin{align*}
\left\vert \psi\right\rangle  &  =\alpha_{1}\left\vert 1\right\rangle
+\cdots+\alpha_{N}\left\vert N\right\rangle ,\\
U\left\vert \psi\right\rangle  &  =\beta_{1}\left\vert 1\right\rangle
+\cdots+\beta_{N}\left\vert N\right\rangle
\end{align*}
are pure states. \ Then consider the network $G$\ shown in Figure
\ref{flowfig}.%
We have a source vertex $s$, a sink vertex $t$, and $N$ input and $N$ output
vertices labeled by basis states $\left\vert 1\right\rangle ,\ldots,\left\vert
N\right\rangle $. \ Each edge of the form $\left(  s,\left\vert i\right\rangle
\right)  $\ has capacity $\left\vert \alpha_{i}\right\vert ^{2}$, each edge
$\left(  \left\vert i\right\rangle ,\left\vert j\right\rangle \right)  $ has
capacity $\left\vert \left(  U\right)  _{ij}\right\vert $, and each edge
$\left(  \left\vert j\right\rangle ,t\right)  $ has capacity $\left\vert
\beta_{j}\right\vert ^{2}$. \ A natural question is how much probability mass
can flow from $s$ to $t$ without violating the capacity constraints.\ \ Rather
surprisingly, we show that one unit of mass (that is, all of it) can.
\ Interestingly, this result would be false if edge $\left(  \left\vert
i\right\rangle ,\left\vert j\right\rangle \right)  $\ had capacity $\left\vert
\left(  U\right)  _{ij}\right\vert ^{2}$\ (or even $\left\vert \left(
U\right)  _{ij}\right\vert ^{1+\varepsilon}$)\ instead of $\left\vert \left(
U\right)  _{ij}\right\vert $. \ We also show that there exists a mapping from
networks to maximum flows in those networks, that is \textit{robust} in the
sense that a small change in edge capacities produces only a small change in
the amount of flow through any edge.

The proofs of these theorems use classical results from the theory of network
flows (see \cite{clrs} for an introduction). \ In particular, let a
\textit{cut} be a set of edges that separates $s$ from $t$; the \textit{value}
of a cut is the sum of the capacities of its edges. \ Then a fundamental
result called the \textit{Max-Flow-Min-Cut Theorem}\ \cite{ff}\ says that the
maximum possible amount of flow from $s$ to $t$ equals the minimum value of
any cut.

\begin{theorem}
\label{flow}One unit of flow can be routed from $s$ to $t$ in $G$.
\end{theorem}

\begin{proof}
By the above, it suffices to show that any cut $C$ in $G$ has value at least
$1$. \ Let $A$ be the set of $i\in\left\{  1,\ldots,N\right\}  $ such
that$\ \left(  s,\left\vert i\right\rangle \right)  \notin C$, and let $B$ be
the set of $j$ such that $\left(  \left\vert j\right\rangle ,t\right)  \notin
C$. \ Then $C$\ must contain every edge $\left(  \left\vert i\right\rangle
,\left\vert j\right\rangle \right)  $\ such that $i\in A$\ and $j\in B$, and
we can assume without loss of generality that $C$ contains no other edges.
\ So the value of $C$ is%
\[
\sum_{i\notin A}\left\vert \alpha_{i}\right\vert ^{2}+\sum_{j\notin
B}\left\vert \beta_{j}\right\vert ^{2}+\sum_{i\in A,~j\in B}\left\vert \left(
U\right)  _{ij}\right\vert .
\]
Therefore we need to prove the matrix inequality%
\[
\left(  1-\sum_{i\in A}\left\vert \alpha_{i}\right\vert ^{2}\right)  +\left(
1-\sum_{j\in B}\left\vert \beta_{j}\right\vert ^{2}\right)  +\sum_{i\in
A,~j\in B}\left\vert \left(  U\right)  _{ij}\right\vert \geq1,
\]
or%
\[
1+\sum_{i\in A,~j\in B}\left\vert \left(  U\right)  _{ij}\right\vert \geq
\sum_{i\in A}\left\vert \alpha_{i}\right\vert ^{2}+\sum_{j\in B}\left\vert
\beta_{j}\right\vert ^{2}.
\]
Let $U$ be fixed, and consider the maximum of the right-hand side over all
$\left\vert \psi\right\rangle $. \ Since%
\[
\beta_{j}=\sum_{i}\left(  U\right)  _{ij}\alpha_{i},
\]
this maximum is equal to the largest eigenvalue $\lambda$\ of the positive
semidefinite matrix%
\[
\sum_{i\in A}\left\vert i\right\rangle \left\langle i\right\vert +\sum_{j\in
B}\left\vert u_{j}\right\rangle \left\langle u_{j}\right\vert
\]
where for each $j$,%
\[
\left\vert u_{j}\right\rangle =\left(  U\right)  _{1j}\left\vert
1\right\rangle +\cdots+\left(  U\right)  _{Nj}\left\vert N\right\rangle .
\]
Let $H_{A}$\ be the subspace of states spanned by $\left\{  \left\vert
i\right\rangle :i\in A\right\}  $, and let $H_{B}$\ be the subspace spanned by
$\left\{  \left\vert u_{j}\right\rangle :j\in B\right\}  $. \ Also, let
$L_{A}\left(  \left\vert \psi\right\rangle \right)  $\ be the length of the
projection of $\left\vert \psi\right\rangle $\ onto $H_{A}$, and let
$L_{B}\left(  \left\vert \psi\right\rangle \right)  $\ be the length of the
projection of $\left\vert \psi\right\rangle $ onto $H_{B}$. \ Then since the
$\left\vert i\right\rangle $'s and $\left\vert u_{j}\right\rangle $'s form
orthogonal bases for $H_{A}$\ and $H_{B}$\ respectively, we have%
\begin{align*}
\lambda &  =\max_{\left\vert \psi\right\rangle }\left(  \sum_{i\in
A}\left\vert \left\langle i|\psi\right\rangle \right\vert ^{2}+\sum_{j\in
B}\left\vert \left\langle u_{j}|\psi\right\rangle \right\vert ^{2}\right) \\
&  =\max_{\left\vert \psi\right\rangle }\left(  L_{A}\left(  \left\vert
\psi\right\rangle \right)  ^{2}+L_{B}\left(  \left\vert \psi\right\rangle
\right)  ^{2}\right)  .
\end{align*}
So letting $\theta$\ be the angle between $H_{A}$\ and $H_{B}$,%
\begin{align*}
\lambda &  =2\cos^{2}\frac{\theta}{2}\\
&  =1+\cos\theta\\
&  \leq1+\max_{\left\vert a\right\rangle \in H_{A},~\left\vert b\right\rangle
\in H_{B}}\left\vert \left\langle a|b\right\rangle \right\vert \\
&  =1+\max_{\substack{\left\vert \gamma_{1}\right\vert ^{2}+\cdots+\left\vert
\gamma_{N}\right\vert ^{2}=1\\\left\vert \delta_{1}\right\vert ^{2}%
+\cdots+\left\vert \delta_{N}\right\vert ^{2}=1}}\left\vert \left(  \sum_{i\in
A}\gamma_{i}\left\langle i\right\vert \right)  \left(  \sum_{j\in B}\delta
_{j}\left\vert u_{j}\right\rangle \right)  \right\vert \\
&  \leq1+\sum_{i\in A,~j\in B}\left\vert \left(  U\right)  _{ij}\right\vert
\end{align*}
which completes the theorem.
\end{proof}

Observe that Theorem \ref{flow}\ still holds if $U$ acts on a mixed state
$\rho$, since we can write $\rho$\ as a convex combination of pure states
$\left\vert \psi\right\rangle \left\langle \psi\right\vert $, construct a flow
for each $\left\vert \psi\right\rangle $\ separately, and then take a convex
combination of the flows.

Using Theorem \ref{flow}, we now define the flow theory $\mathcal{FT}$. \ Let
$F\left(  \rho,U\right)  $ be the set of maximum flows for $\rho
,U$---representable by $N\times N$ arrays of real numbers $f_{ij}$\ such that
$0\leq f_{ij}\leq\left\vert \left(  U\right)  _{ij}\right\vert $\ for all
$i,j$, and also%
\[
\sum_{j}f_{ij}=\left(  \rho\right)  _{ii},~~~\sum_{i}f_{ij}=\left(  U\rho
U^{-1}\right)  _{jj}.
\]
Clearly $F\left(  \rho,U\right)  $ is a convex polytope, which Theorem
\ref{flow}\ asserts is nonempty. \ Form a maximum flow $f^{\ast}\left(
\rho,U\right)  \in F\left(  \rho,U\right)  $ as follows: first let
$f_{11}^{\ast}$\ be the maximum of $f_{11}$\ over all $f\in F\left(
\rho,U\right)  $. \ Then let $f_{12}^{\ast}$\ be the maximum of $f_{12}$\ over
all $f\in F\left(  \rho,U\right)  $\ such that $f_{11}=f_{11}^{\ast}$.
\ Continue to loop through all $i,j$\ pairs in lexicographic order, setting
each $f_{ij}^{\ast}$\ to its maximum possible value consistent with the
$\left(  i-1\right)  N+j-1$\ previous values.

We define the joint probabilities matrix $P$\ by \textquotedblleft
symmetrizing\textquotedblright\ $f^{\ast}\left(  \rho,U\right)  $ over
permutations of basis states---that is,%
\[
P\left(  \rho,U\right)  =\frac{1}{N!}\sum_{Q}\vspace{0pt}Qf^{\ast}\left(
Q^{-1}\rho Q,Q^{-1}UQ\right)  Q^{-1}%
\]
where $Q$ ranges over all $N\times N$\ permutation matrices. \ As discussed in
Section \ref{HV}, given $P$\ we can easily obtain the stochastic matrix $S$ by
dividing the $i^{th}$\ column by $\left(  \rho\right)  _{ii}$, or taking a
limit in case $\left(  \rho\right)  _{ii}=0$. \ It is easy to check that
$\mathcal{FT}$\ so defined satisfies the symmetry and indifference axioms.

Showing that $\mathcal{FT}$\ satisfies robustness is harder. \ Our proof is
based on the Ford-Fulkerson algorithm \cite{ff}, a classic algorithm for
computing maximum flows that works by finding a sequence of \textquotedblleft
augmenting paths,\textquotedblright\ each of which increases the flow from $s$
to $t$ by some positive amount.

\begin{theorem}
\label{fdrobust}$\mathcal{FT}$ satisfies robustness.
\end{theorem}

\begin{proof}
Let $G$ be an arbitrary flow network with source $s$, sink $t$, and directed
edges $e_{1},\ldots,e_{m}$, where each $e_{i}$\ has capacity $c_{i}$\ and
leads from $v_{i}$\ to $w_{i}$. \ It will be convenient to introduce a
fictitious edge $e_{0}$\ from $t$ to $s$ with unlimited capacity; then
maximizing the flow through $G$ is equivalent to maximizing the flow through
$e_{0}$. \ Suppose we produce a new network $\widetilde{G}$ by increasing a
single capacity $c_{i^{\ast}}$ by some $\varepsilon>0$. \ Let $f^{\ast}$\ be
the optimal flow for $G$, obtained by first maximizing the flow $f_{0}%
$\ through $e_{0}$, then maximizing the flow $f_{1}$\ through $e_{1}$ holding
$f_{0}$ fixed, and so on up to $f_{m}$. \ Let $\widetilde{f}^{\ast}$\ be the
maximal flow for $\widetilde{G}$\ produced in the same way. \ We claim that
for all $i\in\left\{  0,\ldots,m\right\}  $,%
\[
\left\vert \widetilde{f}_{i}^{\ast}-f_{i}^{\ast}\right\vert \leq\varepsilon.
\]
To see that the theorem follows from this claim: first, if $f^{\ast}$\ is
robust under adding $\varepsilon$\ to $c_{i^{\ast}}$, then it must also be
robust under subtracting $\varepsilon$\ from $c_{i^{\ast}}$.\ \ Second, if we
change $\rho,U$\ to $\widetilde{\rho},\widetilde{U}$\ such that $\left\Vert
\widetilde{\rho}-\rho\right\Vert \leq1/q\left(  N\right)  $ and $\left\Vert
\widetilde{U}-U\right\Vert \leq1/q\left(  N\right)  $, then we can imagine the
$N^{2}+2N$ edge capacities are changed one by one, so\ that%
\begin{align*}
\left\Vert f^{\ast}\left(  \widetilde{\rho},\widetilde{U}\right)  -f^{\ast
}\left(  \rho,U\right)  \right\Vert  &  \leq\sum_{ij}\left\vert \left\vert
\left(  \widetilde{U}\right)  _{ij}\right\vert -\left\vert \left(  U\right)
_{ij}\right\vert \right\vert +\sum_{i}\left\vert \left(  \widetilde{\rho
}\right)  _{ii}-\left(  \rho\right)  _{ii}\right\vert \\
&  ~~~~~~~~~~~~+\sum_{j}\left\vert \left(  \widetilde{U}\widetilde{\rho
}\widetilde{U}^{-1}\right)  _{jj}-\left(  U\rho U^{-1}\right)  _{jj}%
\right\vert \\
&  \leq\frac{4N^{2}}{q\left(  N\right)  }.
\end{align*}
(Here we have made no attempt to optimize the bound.) \ Third, symmetrizing
over all row and column permutations can only decrease $\left\Vert
\widetilde{P}-P\right\Vert $, not increase it.

We now prove to the claim. \ To do so we describe an iterative algorithm for
computing $f^{\ast}$. \ First maximize the flow $f_{0}$\ through $e_{0}$, by
using the Ford-Fulkerson algorithm \cite{ff}\ to find a maximum flow from $s$
to $t$. \ Let $f^{\left(  0\right)  }$ be the resulting flow, and let
$G^{\left(  1\right)  }$\ be the residual network that corresponds to
$f^{\left(  0\right)  }$. \ For each $i$, that is, $G^{\left(  1\right)  }%
$\ has an edge $e_{i}=\left(  v_{i},w_{i}\right)  $\ of capacity
$c_{i}^{\left(  1\right)  }=c_{i}-f_{i}^{\left(  0\right)  }$, and an edge
$\overline{e}_{i}=\left(  w_{i},v_{i}\right)  $\ of capacity $\overline{c}%
_{i}^{\left(  1\right)  }=f_{i}^{\left(  0\right)  }$.\ \ Next maximize
$f_{1}$\ subject to $f_{0}$ by using the Ford-Fulkerson algorithm to find
\textquotedblleft augmenting cycles\textquotedblright\ from $w_{1}$\ to
$v_{1}$\ and back to $w_{1}$ in $G^{\left(  1\right)  }\setminus\left\{
e_{0},\overline{e}_{0}\right\}  $. \ Continue in this manner until each of
$f_{1},\ldots,f_{m}$\ has been maximized subject to the previous $f_{i}$'s.
\ Finally set $f^{\ast}=f^{\left(  m\right)  }$.

Now, one way to compute $\widetilde{f}^{\ast}$ is to start with $f^{\ast}$,
then repeatedly \textquotedblleft correct\textquotedblright\ it by applying
the same iterative algorithm to maximize $\widetilde{f}_{0}$, then
$\widetilde{f}_{1}$, and so on. \ Let $\varepsilon_{i}=\left\vert
\widetilde{f}_{i}^{\ast}-f_{i}^{\ast}\right\vert $;\ then we need to show that
$\varepsilon_{i}\leq\varepsilon$\ for all $i\in\left\{  0,\ldots,m\right\}  $.
\ The proof is by induction on $i$. \ Clearly $\varepsilon_{0}\leq\varepsilon
$, since increasing $c_{i^{\ast}}$ by $\varepsilon$\ can increase the value of
the minimum cut from $s$ to $t$ by at most $\varepsilon$. \ Likewise, after we
maximize $\widetilde{f}_{0}$,\ the value of the minimum cut from $w_{1}$\ to
$v_{1}$\ can increase by at most $\varepsilon-\varepsilon_{0}+\varepsilon
_{0}=\varepsilon$. \ For of the at most $\varepsilon$\ new units of flow from
$w_{1}$\ to $v_{1}$\ that increasing $c_{i^{\ast}}$\ made available,
$\varepsilon_{0}$\ of them were \textquotedblleft taken up\textquotedblright%
\ in maximizing $\widetilde{f}_{0}$, but the process of maximizing
$\widetilde{f}_{0}$\ could have again increased the minimum cut from $w_{1}%
$\ to $v_{1}$ by up to $\varepsilon_{0}$. \ Continuing in this way,%
\[
\varepsilon_{2}\leq\varepsilon-\varepsilon_{0}+\varepsilon_{0}-\varepsilon
_{1}+\varepsilon_{1}=\varepsilon,
\]
and so on up to $\varepsilon_{m}$. \ This completes the proof.
\end{proof}

That $\mathcal{FT}$\ violates decomposition invariance now follows from
Theorem \ref{decomp}, part (i). \ However, it might be helpful to see an
explicit counterexample. \ Let $I$ be the $1$-qubit maximally mixed state, and
let $R_{\pi/4}$\ be a $\pi/4$\ rotation. \ Then $R_{\pi/4}IR_{\pi/4}^{-1}=I$,
and%
\[
S\left(  I,R_{\pi/4}\right)  =\left[
\begin{array}
[c]{cc}%
1 & 0\\
0 & 1
\end{array}
\right]
\]
since the $1/2$ unit of flow from $\left\vert 0\right\rangle $\ all gets
routed to $\left\vert 0\right\rangle $, and then the $1/2$ unit of flow from
$\left\vert 1\right\rangle $\ can only be routed to $\left\vert 1\right\rangle
$. \ On the other hand, let $\left\vert \varphi_{\theta}\right\rangle
=\cos\theta\left\vert 0\right\rangle +\sin\theta\left\vert 1\right\rangle
$;\ then $S\left(  \left\vert \varphi_{\pi/8}\right\rangle ,R_{\pi/4}\right)
$\ and $S\left(  \left\vert \varphi_{5\pi/8}\right\rangle ,R_{\pi/4}\right)
$\ clearly do \textit{not} equal the identity, since $\cos^{2}\left(
\pi/8\right)  $\ units of flow cannot be routed along an edge of capacity only
$1/\sqrt{2}$. \ Therefore%
\[
S\left(  I,R_{\pi/4}\right)  \neq\frac{S\left(  \left\vert \varphi_{\pi
/8}\right\rangle ,R_{\pi/4}\right)  +S\left(  \left\vert \varphi_{5\pi
/8}\right\rangle ,R_{\pi/4}\right)  }{2}.
\]

Let us also show that $\mathcal{FT}$\ violates product commutativity. \ Let
$\left\vert \psi\right\rangle =\left\vert \varphi_{\pi/4}\right\rangle
\otimes\left\vert \varphi_{-\pi/8}\right\rangle $ be a $2$-qubit initial
state, and let $R_{\pi/4}^{A}$\ and $R_{\pi/4}^{B}$\ be $\pi/4$\ rotations
applied to the first and second qubits respectively. \ Suppose $R_{\pi/4}^{B}%
$\ is applied first to change the second qubit from $\left\vert \varphi
_{-\pi/8}\right\rangle $\ to $\left\vert \varphi_{\pi/8}\right\rangle $.
\ Then one can check that $\frac{1}{2}\cos^{2}\frac{\pi}{8}$\ probability mass
is routed from $\left\vert 0\right\rangle $ to $\left\vert 0\right\rangle $,
and $\frac{1}{2}\sin^{2}\frac{\pi}{8}$ from $\left\vert 1\right\rangle $ to
$\left\vert 1\right\rangle $; the $1/\sqrt{2}$\ edge capacities never come
into play. \ So $S\left(  \left\vert \psi\right\rangle ,R_{\pi/4}^{B}\right)
$\ is the identity, which implies that%
\[
S\left(  R_{\pi/4}^{B}\left\vert \psi\right\rangle ,R_{\pi/4}^{A}\right)
S\left(  \left\vert \psi\right\rangle ,R_{\pi/4}^{B}\right)  =S\left(
\left\vert \varphi_{\pi/4}\right\rangle \otimes\left\vert \varphi_{\pi
/8}\right\rangle ,R_{\pi/4}^{A}\right)  =\left[
\begin{array}
[c]{cccc}%
0 & 0 & 0 & 0\\
0 & 0 & 0 & 0\\
1 & 0 & 1 & 0\\
0 & 1 & 0 & 1
\end{array}
\right]  .
\]
On the other hand, if $R_{\pi/4}^{A}$\ is applied first to change the state to
$\left\vert 1\right\rangle \otimes\left\vert \varphi_{-\pi/8}\right\rangle $,
then when $R_{\pi/4}^{B}$\ is applied next, at most $1/\sqrt{2}$\ of the
$\cos^{2}\frac{\pi}{8}$\ probability mass at $\left\vert 10\right\rangle
$\ can be routed to $\left\vert 10\right\rangle $; the rest must go to
$\left\vert 11\right\rangle $. \ It follows that%
\[
S\left(  R_{\pi/4}^{A}\left\vert \psi\right\rangle ,R_{\pi/4}^{B}\right)
S\left(  \left\vert \psi\right\rangle ,R_{\pi/4}^{A}\right)  \vspace{0pt}\neq
S\left(  R_{\pi/4}^{B}\left\vert \psi\right\rangle ,R_{\pi/4}^{A}\right)
S\left(  \left\vert \psi\right\rangle ,R_{\pi/4}^{B}\right)  .
\]

\subsection{Schr\"{o}dinger Theory\label{SCHROD}}

Our final hidden-variable theory, which we call the \textit{Schr\"{o}dinger
theory} or $\mathcal{ST}$, is the most interesting one mathematically. \ The
idea---to make a matrix into a stochastic matrix via an iterative process of
row and column rescaling---is natural enough that we came upon it
independently, only later learning that it originated in a 1931 paper of
Schr\"{o}dinger\ \cite{schrodinger}. \ Schr\"{o}dinger gave a pair of
functional integral equations that such an iterative process would solve, but
was unable to prove that those equations always have a solution. \ The
existence and uniqueness of a solution were shown under broad conditions by
Nagasawa \cite{nagasawa}, building on earlier work of Fortet \cite{fortet}%
\ and Beurling \cite{beurling}. \ Our goal is to give what (to our knowledge)
is the first self-contained, reasonably accessible presentation of the main
result in this area; and to interpret that result in what we think is the
correct way: as providing one example of a hidden-variable theory, whose
strengths and weaknesses should be directly compared to those of other theories.

Most of the technical difficulties in
\cite{beurling,fortet,nagasawa,schrodinger}\ arise because the stochastic
process being constructed involves continuous time and particle positions.
\ Here we eliminate those difficulties by restricting attention to discrete
time and to finite-dimensional Hilbert spaces. \ We thereby obtain a
generalized version\footnote{In $\left(  r,c\right)  $-scaling, we are given
an invertible real matrix, and the goal is to rescale all rows and columns to
sum to $1$. \ The generalized version is to rescale the rows and columns to
given values (not necessarily $1$).} of a problem that computer scientists
know as $\left(  r,c\right)  $\textit{-scaling of matrices}. \ Sinkhorn
\cite{sinkhorn}\ gave an algorithm for the $\left(  r,c\right)  $-scaling
problem, which was shown to run in polynomial time by Franklin and Lorenz
\cite{fl}\ (see also Linial, Samorodnitsky, and Wigderson \cite{lsw}).

As in the case of the flow theory, given a unitary $U$ acting on a state
$\rho$, the first step is to replace each entry of $U$ by its absolute value,
obtaining the nonnegative matrix $U^{\left(  0\right)  }$\ defined by $\left(
U^{\left(  0\right)  }\right)  _{ij}:=\left\vert \left(  U\right)
_{ij}\right\vert $. \ We then repeatedly tweak $U^{\left(  0\right)  }$\ to
bring it closer to a joint probabilities matrix $P\left(  \rho,U\right)  $.
\ We want to make the $i^{th}$ column of the matrix sum to $\left(
\rho\right)  _{ii}$, and the $j^{th}$ row sum to $\left(  U\rho U^{-1}\right)
_{jj}$ for all $i,j\in\left\{  1,\ldots,N\right\}  $. \ The stochastic matrix
$S\left(  \rho,U\right)  $ is then readily obtained by normalizing each column
to sum to $1$.

The algorithm is iterative. \ For each $t\geq0$\ we obtain\ $U^{\left(
2t+1\right)  }$\ by normalizing each column $i$ of $U^{\left(  2t\right)  }%
$\ to sum to $\left(  \rho\right)  _{ii}$; likewise we obtain $U^{\left(
2t+2\right)  }$\ by normalizing each row $j$ of $U^{\left(  2t+1\right)  }%
$\ to sum to $\left(  U\rho U^{-1}\right)  _{jj}$. \ More formally,%
\begin{align*}
\left(  U^{\left(  2t+1\right)  }\right)  _{ij}  &  =\frac{\left(
\rho\right)  _{ii}}{\sum_{k}\left(  U^{\left(  2t\right)  }\right)  _{ik}%
}\left(  U^{\left(  2t\right)  }\right)  _{ij},\\
\left(  U^{\left(  2t+2\right)  }\right)  _{ij}  &  =\frac{\left(  U\rho
U^{-1}\right)  _{jj}}{\sum_{k}\left(  U^{\left(  2t+1\right)  }\right)  _{kj}%
}\left(  U^{\left(  2t+1\right)  }\right)  _{ij}.
\end{align*}
The crucial fact is that the above iteration converges. \ Our proof will reuse
a result about network flows from Section \ref{FLOW}, in order to define a
nondecreasing \textquotedblleft progress measure\textquotedblright\ based on
Kullback-Leibler distance.

\begin{theorem}
\label{robust}The limit $U^{\left(  \infty\right)  }=\lim_{t\rightarrow\infty
}U^{\left(  t\right)  }$ exists.
\end{theorem}

\begin{proof}
A consequence of Theorem \ref{flow} is that for every $\rho,U$, there exists
an $N\times N$ array of nonnegative real numbers $f_{ij}$ such that

\begin{enumerate}
\item[(1)] $f_{ij}=0$ whenever $\left\vert \left(  U\right)  _{ij}\right\vert
=0$,

\item[(2)] $f_{i1}+\cdots+f_{iN}=\left(  \rho\right)  _{ii}$ for all $i$, and

\item[(3)] $f_{1j}+\cdots+f_{Nj}=\left(  U\rho U^{-1}\right)  _{jj}$\ for all
$j$.
\end{enumerate}

Given any such array, define a progress measure%
\[
Z^{\left(  t\right)  }=%
{\displaystyle\prod\limits_{ij}}
\left(  U^{\left(  t\right)  }\right)  _{ij}^{f_{ij}},
\]
where we adopt the convention $0^{0}=1$. \ We claim that $Z^{\left(
t+1\right)  }\geq Z^{\left(  t\right)  }$\ for all $t\geq1$. \ To see this,
assume without loss of generality that we are on an odd step $2t+1$, and let
$C_{i}^{\left(  2t\right)  }=\sum_{j}\left(  U^{\left(  2t\right)  }\right)
_{ij}$ be the $i^{th}$\ column sum before we normalize it. \ Then%
\begin{align*}
Z^{\left(  2t+1\right)  }  &  =%
{\displaystyle\prod\limits_{ij}}
\left(  U^{\left(  2t+1\right)  }\right)  _{ij}^{f_{ij}}\\
&  =%
{\displaystyle\prod\limits_{ij}}
\left(  \frac{\left(  \rho\right)  _{ii}}{C_{i}^{\left(  2t\right)  }}\left(
U^{\left(  2t\right)  }\right)  _{ij}\right)  ^{f_{ij}}\\
&  =\left(
{\displaystyle\prod\limits_{ij}}
\left(  U^{\left(  2t\right)  }\right)  _{ij}^{f_{ij}}\right)  \left(
{\displaystyle\prod\limits_{i}}
\left(  \frac{\left(  \rho\right)  _{ii}}{C_{i}^{\left(  2t\right)  }}\right)
^{f_{i1}+\cdots+f_{iN}}\right) \\
&  =Z^{\left(  2t\right)  }\cdot%
{\displaystyle\prod\limits_{i}}
\left(  \frac{\left(  \rho\right)  _{ii}}{C_{i}^{\left(  2t\right)  }}\right)
^{\left(  \rho\right)  _{ii}}.
\end{align*}
As a result of the $2t^{th}$\ normalization step, we had $\sum_{i}%
C_{i}^{\left(  2t\right)  }=1$. \ Subject to that constraint, the maximum of%
\[%
{\displaystyle\prod\limits_{i}}
\left(  C_{i}^{\left(  2t\right)  }\right)  ^{\left(  \rho\right)  _{ii}}%
\]
over the $C_{i}^{\left(  2t\right)  }$'s\ occurs when $C_{i}^{\left(
2t\right)  }=\left(  \rho\right)  _{ii}$\ for all $i$---a simple calculus fact
that follows from the nonnegativity of Kullback-Leibler distance. \ This
implies that $Z^{\left(  2t+1\right)  }\geq Z^{\left(  2t\right)  }$.
\ Similarly, normalizing rows leads to $Z^{\left(  2t+2\right)  }\geq
Z^{\left(  2t+1\right)  }$. \ 

It follows that the limit $U^{\left(  \infty\right)  }=\lim_{t\rightarrow
\infty}U^{\left(  t\right)  }$\ exists. \ For suppose not; then some
$C_{i}^{\left(  t\right)  }$\ is bounded away from $\left(  \rho\right)
_{ii}$, so there exists an $\varepsilon>0$\ such that $Z^{\left(  t+1\right)
}\geq\left(  1+\varepsilon\right)  Z^{\left(  t\right)  }$\ for all even
$t$.\ \ But this is a contradiction, since $Z^{\left(  0\right)  }>0$\ and
$Z^{\left(  t\right)  }\leq1$ for all $t$.
\end{proof}

It is immediate that $\mathcal{ST}$\ satisfies symmetry and indifference.
\ Let us show that it satisfies product commutativity as well.

\begin{proposition}
\label{sdprod}$\mathcal{ST}$ satisfies product commutativity.
\end{proposition}

\begin{proof}
Given a state $\left\vert \psi\right\rangle =\left\vert \psi_{A}\right\rangle
\otimes\left\vert \psi_{B}\right\rangle $,\ let $U_{A}\otimes I$\ act only on
$\left\vert \psi_{A}\right\rangle $\ and let $I\otimes U_{B}$\ act only on
$\left\vert \psi_{B}\right\rangle $. \ Then we claim that%
\[
S\left(  \left\vert \psi\right\rangle ,U_{A}\otimes I\right)  =S\left(
\left\vert \psi_{A}\right\rangle ,U_{A}\right)  \otimes I.
\]
The reason is simply that multiplying all amplitudes in $\left\vert \psi
_{A}\right\rangle $ and $U_{A}\left\vert \psi_{A}\right\rangle $ by a constant
factor $\alpha_{x}$, as we do for each basis state $\left\vert x\right\rangle
$ of $\left\vert \psi_{B}\right\rangle $, has no effect on the scaling
procedure that produces $S\left(  \left\vert \psi_{A}\right\rangle
,U_{A}\right)  $. \ Similarly%
\[
S\left(  \left\vert \psi\right\rangle ,I\otimes U_{B}\right)  =I\otimes
S\left(  \left\vert \psi_{B}\right\rangle ,U_{B}\right)  .
\]
It follows that%
\begin{align*}
S\left(  \left\vert \psi_{A}\right\rangle ,U_{A}\right)  \otimes S\left(
\left\vert \psi_{B}\right\rangle ,U_{B}\right)   &  =S\left(  U_{A}\left\vert
\psi_{A}\right\rangle \otimes\left\vert \psi_{B}\right\rangle ,I\otimes
U_{B}\right)  S\left(  \left\vert \psi\right\rangle ,U_{A}\otimes I\right)
\vspace{0pt}\\
&  =S\left(  \left\vert \psi_{A}\right\rangle \otimes U_{B}\left\vert \psi
_{B}\right\rangle ,U_{A}\otimes I\right)  S\left(  \left\vert \psi
\right\rangle ,I\otimes U_{B}\right)  .
\end{align*}

\end{proof}

On the other hand, let us show that $\mathcal{ST}$\ violates decomposition
invariance. \ Using the same notation as in Section \ref{FLOW}, we have
$R_{\pi/8}IR_{\pi/8}^{-1}=I$, $R_{\pi/8}\left\vert \varphi_{\pi/8}%
\right\rangle =\left\vert \varphi_{\pi/4}\right\rangle $, and $R_{\pi
/8}\left\vert \varphi_{5\pi/8}\right\rangle =\left\vert \varphi_{3\pi
/4}\right\rangle $, from which it can be calculated that%
\begin{align*}
S\left(  I,R_{\pi/8}\right)   &  \approx\left[
\begin{array}
[c]{cc}%
0.707 & 0.293\\
0.293 & 0.707
\end{array}
\right]  ,\\
S\left(  \left\vert \varphi_{\pi/8}\right\rangle ,R_{\pi/8}\right)   &
\approx\left[
\begin{array}
[c]{cc}%
0.555 & 0.177\\
0.445 & 0.823
\end{array}
\right]  ,\\
S\left(  \left\vert \varphi_{5\pi/8}\right\rangle ,R_{\pi/8}\right)   &
\approx\left[
\begin{array}
[c]{cc}%
0.177 & 0.555\\
0.823 & 0.445
\end{array}
\right]  .
\end{align*}
Hence%
\[
S\left(  I,R_{\pi/8}\right)  \neq\frac{S\left(  \left\vert \varphi_{\pi
/8}\right\rangle ,R_{\pi/8}\right)  +S\left(  \left\vert \varphi_{5\pi
/8}\right\rangle ,R_{\pi/8}\right)  }{2}.
\]

\section{Discussion\label{DISC}}

The idea that certain observables in quantum mechanics might have trajectories
governed by dynamical laws has reappeared many times: in Schr\"{o}dinger's
1931 stochastic approach \cite{schrodinger}, Bohmian mechanics \cite{bohm},
modal interpretations \cite{bd,dickson,dieks}, and elsewhere. \ Yet because
all of these proposals yield the same predictions for single-time
probabilities, if we are to decide between them it must be on the basis of
internal mathematical considerations. \ A main message of this paper has been
that such considerations can actually get us quite far.

To focus attention on the core issues, we restricted attention to the simplest
possible setting: discrete time, a finite-dimensional Hilbert space, and a
single orthogonal basis. \ Within this setting, we proposed what seem like
reasonable axioms that any hidden-variable theory should satisfy: for example,
symmetry under permutation of basis states, robustness to small perturbations,
and independence of the temporal order of spacelike-separated events. \ We
then showed that not all of these axioms can be satisfied simultaneously.
\ But perhaps more surprisingly, we also showed that certain subsets of axioms
\textit{can} be satisfied for highly nontrivial reasons. \ In showing that the
indifference and robustness axioms can be simultaneously satisfied, Section
\ref{SPECIFIC}\ revealed an unexpected connection between unitary matrices and
the classical theory of network flows.

As mentioned previously, the most important open problem is to show that the
Schr\"{o}dinger theory\ satisfies robustness. \ Currently, we can only show
that the matrix $P_{\mathcal{ST}}\left(  \rho,U\right)  $\ is robust to
\textit{exponentially} small perturbations, not polynomially small ones. \ The
problem is that if any row or column sum in the $U^{\left(  t\right)  }%
$\ matrix is extremely small, then the $\left(  r,c\right)  $-scaling process
will magnify tiny errors in the entries. \ Intuitively, though, this effect
should be washed out by later scaling steps.

A second open problem is whether there exists a theory that satisfies
indifference, as well as commutativity for all separable \textit{mixed} states
(not just separable pure states). \ A third problem is to investigate other
notions of robustness---for example, robustness to small
\textit{multiplicative} rather than additive errors.

\section{Acknowledgments}

I thank Umesh Vazirani, Ronald de Wolf, and an anonymous reviewer for comments
on an earlier version of this paper; Dorit Aharonov, Guido Bacciagaluppi, John
Preskill, and Avi Wigderson for helpful discussions; and Dennis Dieks for correspondence.

\end{document}